\documentclass[12pt]{article}

\usepackage{amssymb,amsmath}
\usepackage{graphicx}
\usepackage{hyperref}
\usepackage{xcolor}
\usepackage[normalem]{ulem}

\title{Scattering of knotted vortices (Hopfions)\\ in the Faddeev-Skyrme
  model}

\author { J.  Hietarinta${}^{1}\footnote{corresponding author, e-mail:
    hietarin@utu.fi}$, J. Palmu${}^1$,
  J. J\"aykk\"a${}^{1,2}$ and P. Pakkanen${}^1$ \\
\\
\normalsize{$^{1}$ Department of Physics and Astronomy,
  University of Turku,}\\
\normalsize{  FI-20014 Turku, Finland}
\\
\normalsize{  ${}^2$ School of Mathematics
  University of Leeds,}\\
\normalsize{  LS2 9JT Leeds, United Kingdom}\\
}

\date{}

\begin{document} 

\maketitle 

\baselineskip18pt

\begin{abstract}
  Several materials, such as ferromagnets, spinor Bose-Einstein
  condensates, and some topological insulators, are now believed to
  support knotted structures. One of the most successful
  base-models having stable knots is the Faddeev-Skyrme model and it
  is expected to be contained in some of these experimentally relevant
  models.  The taxonomy of knotted topological solitons (Hopfions) of
  this model is known.  In this paper we describe some aspects of the
  dynamics of Hopfions and show that they do indeed behave like
  particles: during scattering the Hopf charge is conserved and bound
  states are formed when the dynamics allows it. We have also
  investigated the dynamical stability of a pair of Hopfions in
  stacked or side-by-side configurations, whose theoretical stability
  has been recently discussed by Ward.
\end{abstract}

\section{Introduction}
\subsection{Background}
Topological solitons play an important role in many areas of
physics. They can be beneficial, like the Abrikosov vortices in type
II superconductors or unwanted like the dislocations in nematic liquid
crystals. Likewise, the creation and observation of topological
solitons in Bose-Einstein condensates is nowadays routine in
laboratories around the world.  Recent experimental observations have
renewed interest in the types of models that can support topological
solitons, and it is important to understand the basic properties and
phenomenology of topological solitons in these models.  Theoretically,
the possibility of the presence of skyrmion structures in certain
materials has been known for long\cite{1993PhRvB..4716419S,
  1995JETPL..62..247B} (for a more recent work and references, see
\cite{2006Natur.442..797R}), and this has recently
been experimentally verified at least in
MnSi\cite{2009Sci...323..915M} and Fe$_{1-x}$Co$_{x}$Si
\cite{2010Natur.465..901Y}. A newer theoretical discovery is the
family of materials collectively known as topological insulators (for
a review, see \cite{Moo10}). The experimental observation of
topological insulator phase in Bi$_{1-x}$Sb$_x$ \cite{D.Hsieh02132009},
Bi$_2$Se$_3$\cite{2009NatPh...5..398X} and
Bi$_2$Te$_3$\cite{2009Sci...325..178C} confirmed these theoretical
predictions and served to further add relevance to the basic research
involving topological solitons. For the present work, perhaps the most
interesting possibility is the discovery that frustrated magnetic
materials may support topological insulator phases, where the
wave-function in momentum-space is classified by the Hopf
invariant\cite{Moore:2008aa}. As will be discussed below, when the
system is described by the Hopf charge, it will support knotted
structures.

Knotted structures have a long history in physics. They were first
considered by Lord Kelvin, who proposed in 1867 \cite{Kelvin} that
atoms could be knotted tubes of ether. This idea did not yield a
satisfactory atomic theory, but subsequently more realistic models
have been proposed with potential for knotted structures. This has
been done, for example, in the context of ferromagnets\cite{SutPRB07},
Bose-Einstein condensates\cite{KNU08}, and optics\cite{DKJOP:Nat10}. A
unifying feature of all these, in addition to the knotted structures,
is that all these phenomena and their knots can be described by
classical field theory.

\subsection{Hopf charge}

In a field theoretical description the knot is not made of a rope in
empty space but through the twisting of a globally defined vector
field. Here we will only consider the case in which the knottedness is
characterized by the Hopf charge.  The minimal model with stable
knotted structures seems to be the Faddeev-Skyrme model
\cite{Faddeev:1975} and it is believed that this model is contained
within various models with immediate physical application, such as the
two-component Ginzburg-Landau model modified with an additional Ward
like term.  \cite{Babaev:2001zy}.

The physical carrier field in the Faddeev-Skyrme model is a smooth 3D
unit vector field $\phi=(\phi_1,\phi_2,\phi_3)$ with ${\phi({\bf
    x})}\cdot {\phi({\bf x})}=1$. Unit vector fields with such
properties have been proposed to exist, e.g., in super-fluid
${}^{3}$He in its A-phase (Continuous Unlocked
Vortex)\cite{Parts:1995} and in ferromagnets\cite{SutPRB07}. In order
to be able to define the Hopf charge it is also necessary that the
vector field $\phi$ approaches the same value at all asymptotic
directions: $\phi({\bf x})\to \phi_\infty,$ when $ |{\bf x}|\to\infty$
(asymptotic triviality). Asymptotic triviality means that from the
point of view of $\phi$ the Euclidean space ${R}^3$ is topologically
like the sphere $S^3$ and this allows one to define the Hopf charge as
an element of the homotopy class $\pi_3(S^2)={\mathbb Z}$.  A concrete Hopfion
vortex ring with Hopf charge $Q=-1$ is given by
\begin{equation}\label{q1n}
{\phi}=
\left(\frac{4(2{x}z-{y}(r^2-1))}{(1+r^2)^2},
  \frac{4(2{y}z+{x}(r^2-1))}{(1+r^2)^2},
  \frac{8(r^2-z^2)}{(1+r^2)^2}-1\right),
\end{equation}
where $r^2=x^2+y^2+z^2$, however, this configuration is not a solution
for the equation of motion studied here. It is easy to see that
${\phi}=(0,0,-1)$ at infinity in any direction and also on the
$z$-axis, while the preimage of ${\phi}=(0,0,+1)$, which is defined as
the vortex core, is the ring $x^2+y^2=1,\,z=0$.

\begin{figure}[f]
\begin{center}
  \includegraphics[width=6cm]{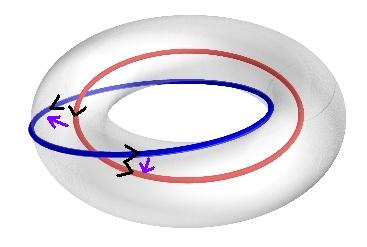}
\end{center}
\caption{Red line: the preimage of ${\phi}=(0,0,+1)$ (vortex core);
  blue line: the preimage of ${\phi}=(\frac34,0,\frac45)$. The lines
  have been assigned compatible directions and their linking number
  can then be computed as explained in text.  \label{F:apu}}
\end{figure}

Since the core is just a simple loop it does not explain the Hopf
charge. For that purpose we need to consider also the preimage-curve
of some other value of $\phi$ and the linking of these two curves. Let
us for example consider the tubular iso-surface defined by
$\phi_3=\frac45$ and on it the curve that is the preimage of
${\phi}=(\frac35,0,\frac45)$, see Figure \ref{F:apu}.  The linking
number of this curve with the core-line can now be computed using the
standard rules of knot theory: If the top arrow can be aligned with
the bottom arrow using a clockwise rotation, the crossing is assigned
$-1$, for a counterclockwise rotation +1.  The linking number is then
$1/2$ of the sum of signed crossings. From Figure \ref{F:apu} we can
see that there are two crossings of signature $-1$ and therefore the
Hopf charge is $Q=-1$.

The Hopf charge can also be computed by integrating a charge density:
For the antisymmetric field $F_{ij}:={\phi\cdot\partial_i
  \phi\times\partial_j \phi}$ one constructs the potential $A_i$ such that
$F_{ij}=\partial_i A_j- \partial_j A_i$ and then the Hopf charge is
given by
\begin{equation}
Q=\frac1{32\pi^2}\int \epsilon^{ijk} A_i F_{jk}\, d^3x.\label{Q}
\end{equation}
A proof that these two methods give the same result is given in
\cite{TB}.

\subsection{The dynamics}
For the dynamics we use a variation of the Skyrme model as proposed by
Faddeev in 1975\cite{Faddeev:1975}, it is defined by the Lagrangian
\begin{equation}\label{lag}
L=\int\left[(\partial_\mu {\phi})^2 + {g}\,
F_{\mu\nu}^2\right] d^3x,\quad
F_{\mu\nu}:={\phi\cdot\partial_\mu \phi\times\partial_\nu \phi}.
\end{equation}
Note that in the static case we have the following scaling property:
under the scaling $r\to\lambda r$ the integrated kinetic term scales
as ${\lambda}$ while the integrated $F^2$ term scales as
${\lambda^{-1}}$.  Thus according to the virial theorem nontrivial
configurations will attain some fixed size determined by the {\em
  dimensional} coupling constant {$g$}. In the computations reported
below we used $g=1/2$.

A very important property of this model is that the energy is bounded
from below\cite{Vakulenko_Kapitanskii:1979} and from
above\cite{LinYang2004}
\begin{equation}\label{VK}
h'\, |Q|^{\frac34}\le E\le h\, |Q|^{\frac34},
\end{equation}
where $h,h'$ are some constants and $Q$ the Hopf charge.  Since the
exponent of the charge is less than 1 it follows that configurations with higher
Hopf charge tend to form bound states rather than split into several
charge 1 states. 

An intriguing question is the form of the minimum energy states with
higher Hopf charge. After some tentative
works\cite{Faddeev:1997zj,Gladikowski:1997mb} detailed taxonomy has
been obtained by Battye and
Sutcliffe\cite{Battye:1998zn,Battye:1998pe,Sutcliffe:2007} and
Hie\-ta\-rin\-ta and Salo\cite{Hietarinta:1998kt,Hietarinta:2000ci}.
The results follow very well\cite{Hietarinta:2000ci,Sutcliffe:2007}
the bound (\ref{VK}).  It turns out that a twisted torus-like
configuration (similar to (\ref{q1n})) is a minimum energy state only
for $|Q|\le4$ \cite{Hietarinta:2000ci}, for higher charges linked or knotted
configurations are found. The trefoil knot is persistently found for
charge $|Q|=7$ from a wide variety of initial states, even for long
vortices.\cite{Hietarinta:2003vn}

We will next give the computational method and then describe the
results for two dynamical studies: scattering head on collisions, and
attractive and repulsive channel simulations inspired by the work of
Ward in \cite{Ward2000}.

\section{Computational method}
For scattering dynamics we used the Lagrangian (\ref{lag}) in the full
3+1-dimensional Minkowski space-time (with $c=1$) and the resulting
Euler-Lagrange equations, derived with a Lagrange multiplier to
guarantee $\phi\cdot\phi=1$ during evolution. The second order (in
time) equation was split into two first order equations for numerical
computations. The spatial derivatives are discretized on $5^3$ points.
The formulae are quite similar to those in \cite{Battye:2001qn}. Time
evolution was computed using 4th order Runge-Kutta (five stage low
memory version\cite{RK}). For scattering studies the computational
lattice varied between $200^3$ and $300^3$, but all the simulations used
the same lattice constant $h=0.15$. In the three-Hopfion collisions
the physical size of the grid was $45.0 \times 30.0\times 30.0$ and in
the double-collisions $30.0^3$. The constant of
Courant-Friedrichs-Lewy like condition linking time and space
discretization is $c_{fl} \equiv \Delta t/h = 0.02$. Ward's
channel simulations were computed in a lattice of $300^3$ unless
otherwise noted; the grid constant was still $h=0.15$ but
$c_{fl}=0.2$.  The Courant instability was controlled by smoothing.
For computational platform we used SAMRAI\cite{samrai}.

As our work progressed the significance of boundary condition became
more and more clear and thus we have tried a variety of different
methods and parameter configurations to minimize boundary errors. The
presented two- and three-Hopfions scatterings are calculated with
Dirichlet boundary conditions
${\phi}|_\partial={\phi}_\infty=(0,0,-1)$, and the Ward's channel
simulations with Neumann boundary condition
$\partial_{\{x,y,z\}}{\phi}|_\partial=(0,0,0)$. Furthermore, in the two-
and three-Hopfion scatterings we had a thin and mild absorption layer
near the boundary in order to prevent radiation from reflecting and
re-entering the scattering region.

Visualization of a vector field is problematic. Since we cannot plot
the vector field at all points of the 3D space we chose to plot
iso-surfaces determined by the value of the third component $\phi_3$,
often with $\phi_3=0.7$, which is a narrow tube around the core.  This
tube is painted\cite{Hietarinta:1998kt} with color determined by the
first two components of $\phi$. For visualization platform we used
VisIt\cite{visit}.

\section{Results}
\subsection{The fundamental deformation processes}
Before analyzing specific scattering processes it is important to
recall the possible elementary deformation processes. It was already
noted above that it is not enough to consider only the core-line but
also how the unit-vector field twists around it. In fact, the proper
knot theoretical setting is to use {\em framed links} which can be
realized as {\em directed ribbons} (pre-images of line
segments\cite{HJS:2002}). This adds local information near the curve,
e.g., twisting around it.  Knot deformations correspond to ribbon
deformations, which allow certain types of crossing and breaking, in
which the Hopf charge will nevertheless be conserved. Examples of
ribbon deformations can be seen in\cite{HJS:2002}.

\subsection{Collisions and scattering}
\label{Collisions and scattering}
We will now consider the scattering of low-charge ``un-knot'' Hopfions
in two different situations: three-body scattering on a fixed target
with projectiles from left and right, and two-body scattering with
impact parameter. The the individual Hopfions (minimum energy states)
were first created using the steepest descent method (corresponding to
1st order dissipative dynamics with a fictitious time)
\cite{Hietarinta:1998kt}.  These Hopfions were then rotated, Lorentz
boosted and combined to form the initial configurations for
scattering. This was done using interpolation routines provided by the
SciPy python library.

\begin{figure}[f]
\begin{center}
  \includegraphics[width=8.8cm]{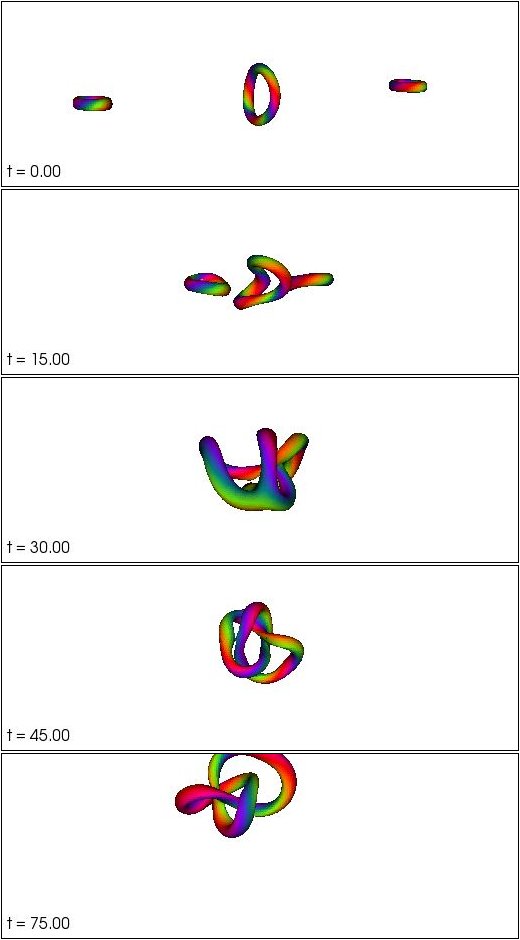}
\end{center}
\caption{The formation of a trefoil knot by the scattering of two
  $Q=-2$ Hopfions with velocities $\pm0.75$ on a stationary $Q=-3$
  target. For more details see
  \href{http://www.youtube.com/watch?v=zkHCtrP6S_Y}{Movie
    S1}.\label{F1}}
\end{figure}

In the first scattering process, for which snapshots are given in Figure
\ref{F1}, with full details in
\href{http://www.youtube.com/watch?v=zkHCtrP6S_Y}{Movie S1}, we have
two charge $Q=-2$ Hopfions approaching a stationary $Q=-3$ target head
on, from left and right. The moving Hopfions were Lorentz boosted
numerically to speeds $0.75$ and $-0.75$, respectively, and then
immersed into the computational box.  Since the total charge is $Q=-7$
we expect the result to be a trefoil knot. This is indeed the case,
after several intermediate steps. The resulting trefoil is not
stationary but rather oscillates and rotates, and eventually the
Hopfion settles down drifting slowly.  Since $7^{3/4}/(2\cdot
2^{3/4}+3^{3/4})\approx 0.76$ there is theoretically 24\% excess
energy, some of which will radiate as low amplitude, close to vacuum
($\phi_3\approx -1$), waves from the scattering region.
These are not visible at the iso-surface $\phi_3=0.7$ level. The
resulting drift is probably due to momentum conservation in an
asymmetric radiation background.

\begin{figure}[f]
\begin{center}
  \includegraphics[width=8.8cm]{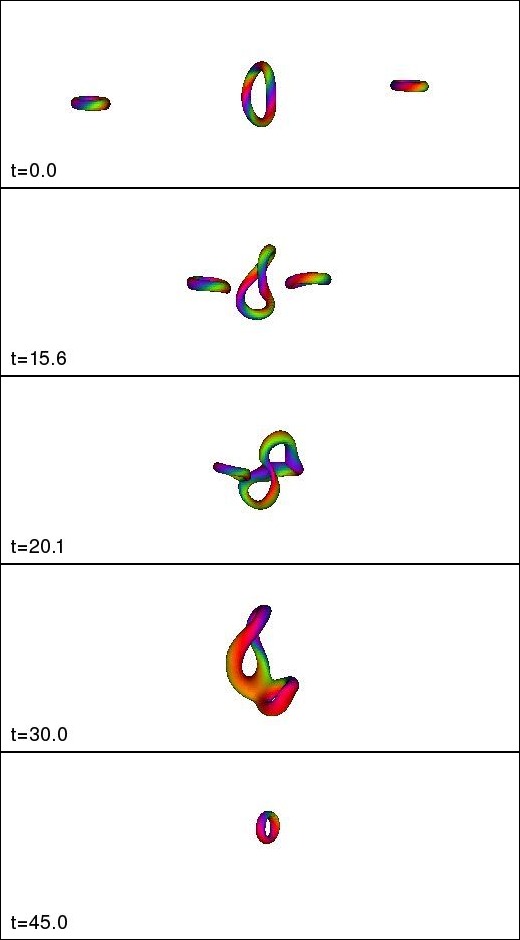}
\end{center}
\caption{Collision of three Hopfions of charges $Q=-2,+3,-2$ and
  velocities $0.75,\, 0,\, -0.75$, respectively, resulting
  with a Hopfion of charge $-1$. For details see
  \href{http://www.youtube.com/watch?v=3x2Ns8AkBLE}{Movie S2}.
  \label{F4}}
\end{figure}

\begin{figure}[f]
\begin{center}
  \includegraphics[width=8.8cm]{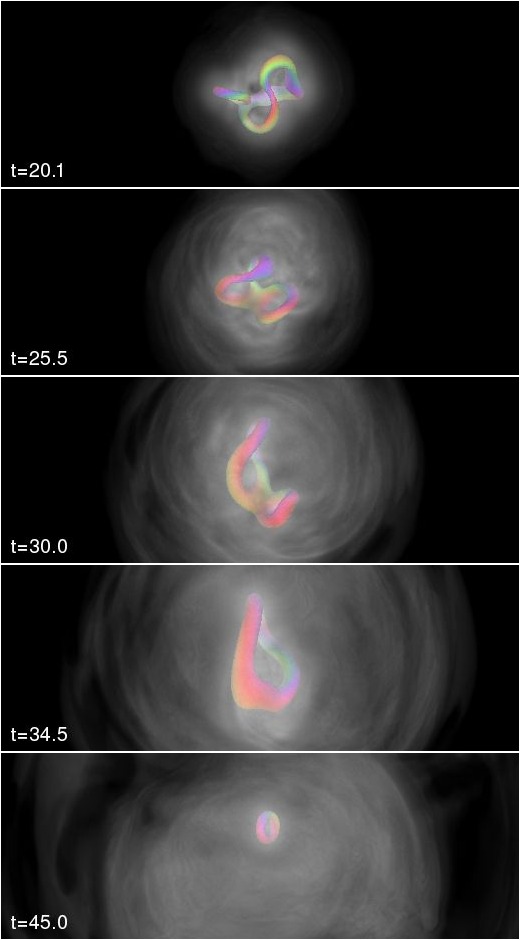}
\end{center}
\caption{Same scattering process as in Figure \ref{F4} but with
  further intermediate states for the later stage of the deformation
  process. Here we have added the energy density 
  plot which appears as a grayish halo. See also
  \href{http://www.youtube.com/watch?v=ywWlA0sDAqo}{Movie S3}.
  \label{F5}}
\end{figure}

In Figures \ref{F4},\ref{F5} (for details see
\href{http://www.youtube.com/watch?v=3x2Ns8AkBLE}{Movies S2} and
\href{http://www.youtube.com/watch?v=ywWlA0sDAqo}{S3}) we have a
situation similar to the one in Figure \ref{F1}, except that now the
central target has charge $Q=+3$. The total charge is then $Q=-1$ and
indeed the final state has one Hopfion of $Q=-1$. As the Hopfions
approach each other they deform by the rule ``like colors attract'',
thereby preparing for ribbon crossings.  Since $1/(2\cdot
2^{3/4}+3^{3/4})\approx 0.18$ most of the energy has to radiate away.
This can be clearly seen in Figure \ref{F5}
(\href{http://www.youtube.com/watch?v=ywWlA0sDAqo}{Movie S3}) which
describes in further detail the later stages of the process.  In this
figure we have indicated the distribution of energy density by a white
cloud. Up to $t=20.1$ the energy is localized near the vortex tube but
when the deformation towards $Q=-1$ continues more radiation is
released and can be seen to radiate away.

\begin{figure}[f]
\begin{center}
  \includegraphics[width=8.8cm]{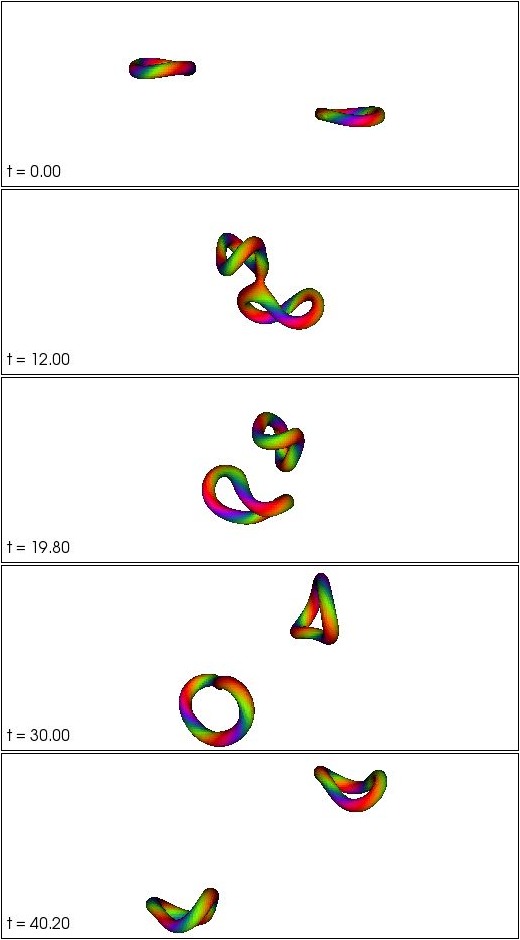}
\end{center}
\caption{Collision of two Hopfions of charge $Q=-3$. The initial
  velocities were $\pm 0.5$ and the impact parameter $3.75$. For more
  details see \href{http://www.youtube.com/watch?v=nRw1z81i4hQ}{Movie
    S4}. \label{F2}}
\end{figure}

For the two-body scattering with impact parameter we present the
scattering of two $Q=-3$ Hopfions with two different impact
parameters.  In Figure \ref{F2}
\href{http://www.youtube.com/watch?v=nRw1z81i4hQ}{(Movie S4)} we have
the impact parameter $3.75$ and the Hopfions have speed $0.5$.
Although the Hopfions touch briefly at $t=12$ the speed and distance
prevent a bound state from forming. The individual Hopfions just
continue tumbling along a slightly bent trajectory.

\begin{figure}[f]
\begin{center}
  \includegraphics[width=8.8cm]{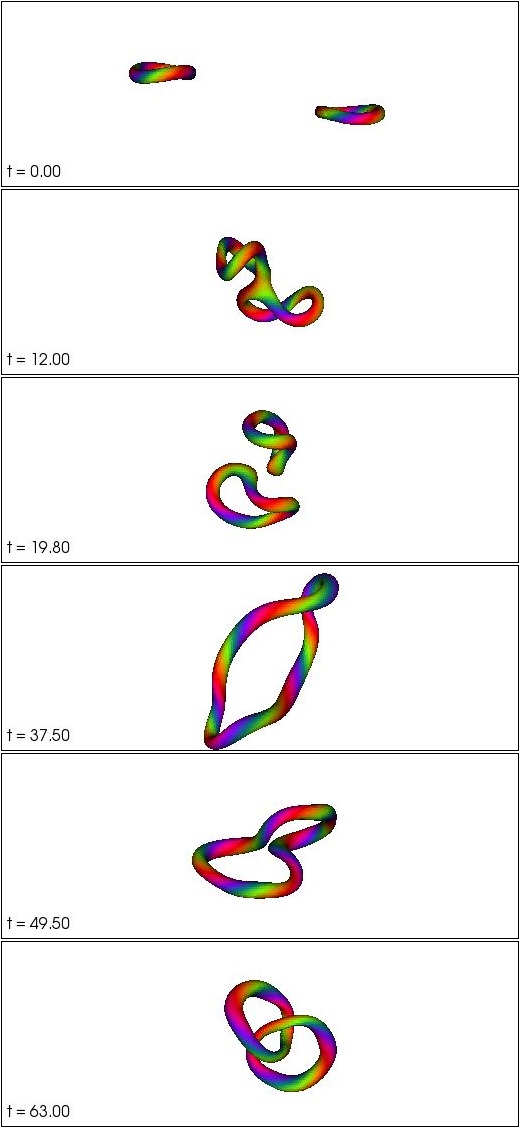}
\end{center}
\caption{Same as in Figure \ref{F2} except that impact parameter is
  $3.6$. For more details see
  \href{http://www.youtube.com/watch?v=JqIDQwAQjOw}{Movie
    S5}. \label{F3}}
\end{figure}

For the same speeds but with slightly smaller impact parameter value
$3.6$ the result is entirely different (Figure \ref{F3},
\href{http://www.youtube.com/watch?v=JqIDQwAQjOw}{Movie S5}). For a
moment it seems that the Hopfions would again continue along their
individual trajectories but there is just enough time to form an
elongated loop. After this the evolution is typical for the total
charge $|Q|=6$ case: The loop behaves like an over-twisted rubber band
and proceeds to make one ribbon crossing deformation to reach the
linked loop configuration that is standard for this Hopf charge. The
linked configuration continues to rotate and oscillate.

The above results illustrate the difficulty associated with trying to
make any generalizations about the behavior of Hopfions. However, some
progress has been made by modeling Hopfions with elastic
rods\cite{2011PhRvD..83f5008H}. It would be interesting to see if the
elastic rod model could be extended to approximate dynamical features
as well.

\subsection{Static initial states}
\label{Static initial states}

We will next consider the case of two unit charge Hopfion unknots
placed on static positions not too far from each other. This case was
studied by Ward \cite{Ward2000}, who observed (using energy
minimization) that under symmetry preserving relaxation an
axisymmetric two Hopfion state reaches a local minimum ($E=2.26$)
where the soliton cores around the symmetry axis are separated by a
nonzero but finite distance. He also noticed that this separated
configuration has $13 \%$ higher total energy than the twice wound
multisoliton state ($E=2.00$) both of which are in the same homotopy
class. The nature of the axisymmetric state is not yet clear and, as
Ward stated, more study is needed to investigate whether the state is
actually a local minimum or just a saddle-point.

For these cases we prepared the initial states using Wards proposal,
that in the stereographic coordinates $W=(\phi_1-i\phi_2)/(1+\phi_3)$,
a good ansatz for a $|Q|=1$ Hopfion is given by $ W=(x+iy)/(z-if(r))$,
where $r=(x^2+y^2+z^2)^{1/2}$ and $f$ is some specific function of $r$
(see Equation (4) in \cite{Ward2000}, Ward uses
$W=(\phi_1+i\phi_2)/(1+\phi_3)$).  Actually this can be generalized by
adding some signs in the expression:
\begin{equation}\label{WA}
W=\frac{\sigma_1x+i\sigma_2y}{z-i\tau\sigma_1\sigma_2f(r)}.
\end{equation}
Here the sign of $\tau$ determines whether this is a Hopfion or
anti-Hopfion, while the signs of the $\sigma$'s can be changed
by $180^o$ rotations around one of the coordinate axes.
Thus for a single Hopfion the signs $\sigma_i$ are irrelevant but they
do play a role when we construct an initial state with two Hopfions.
\footnote{Note that in the
  asymptotic limit the situation is simpler because from the
  expression
\begin{equation*}
W=\frac{\sigma_1x+i\sigma_2y}{-i\tau\sigma_1\sigma_2f(r_a)}
=\frac{x+i\sigma_1\sigma_2y}{-i\tau\sigma_2f(r_a)}
\end{equation*}
one cannot extract $\tau$ in order to say whether it describes a
Hopfion or an anti-Hopfion.}

\subsubsection{Stacked Hopfions}
\label{Stacked Hopfions}
It was noted by Ward \cite{Ward2000} that a good two Hopfion ansatz is
obtained just by adding the corresponding $W$ functions \eqref{WA}.
Thus for the case with two Hopfions located at  $z=\pm a$ and sharing
the same axis (Ward's Channel A) we can take
\begin{equation}\label{2solz}
W=\frac{\sigma_1x+i\sigma_2y}{z-a-i\tau\sigma_1\sigma_2f(r_{z-a})}+
\frac{\sigma_1'x+i\sigma_2'y}{z+a-i\tau\sigma_1'\sigma_2'f(r_{z+a})},
\end{equation}
as the starting configuration.  Here
$r_{z\pm a}=(x^2+y^2+(z\pm a)^2)^{1/2}$.

Ward uses a dipole pair to describe the configuration, in our notation
$P_+=(\sigma_1,0,0),\,Q_+=(0,\sigma_2,0)$ and
$P_-=(\sigma_1',0,0),\,Q_-=(0,\sigma_2',0)$, so that $D_+:=P_+\times
Q_+=(0,0,\sigma_1\sigma_2)$, etc. Furthermore the angle between $P_+$
and $P_-$ is 0 if $\sigma_1=\sigma_1'$ and $180^o$ if
$\sigma_1=-\sigma_1'$.

The signs in \eqref{2solz} can be changed by various $180^o$ rotations
and this way the 16 possible sign combinations
$(\sigma_1,\sigma_2,\sigma_1',\sigma_2')$ can be reduced to a
canonical 6. They are 1) $(++++)$ ($D$'s to same direction, angle
$0^o$), 2) $(++--)$ ($D$'s to same direction, angle $180^o$), 3)
$(+++-)$ ($D$'s away from each other, angle $0^o$), 4) $(++-+)$ ($D$'s
away from each other, angle $180^o$), 5) $(+-++)$ ($D$'s towards each
other, angle $0^o$), 6) $(-+++)$ ($D$'s towards each other, angle
$180^o$).

\begin{figure}[f]
\begin{center}
  \includegraphics[width=12cm]{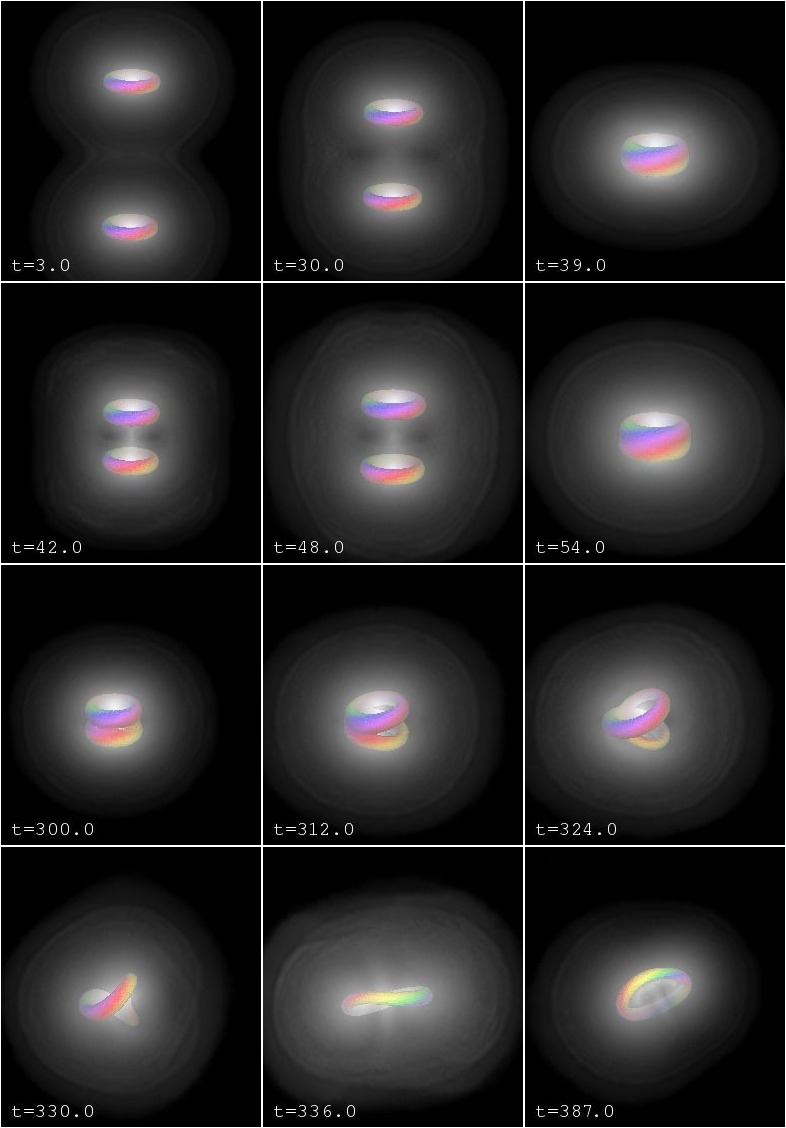}
\end{center}
\caption{Stacked configuration $(++++)$ produces an attractive
  interaction between the two $H=1$ Hopfions. The pair oscillates
  multiple times before a slight asymmetry brings the system to the
  more stable $H=2$ single-ring configuration. For details see
  \href{http://www.youtube.com/watch?v=ERzEU3v0wAk}{Movie W1}, and the
  discussion in the end of section \ref{Stacked Hopfions}.
\label{F7}}
\end{figure}

The time evolution for these cases is as follows. For Cases 1 and 2
the system is in the attractive channel (Figure \ref{F7},
\href{http://www.youtube.com/watch?v=ERzEU3v0wAk}{Movie W1}).  The
Hopfions approach each other, bounce a few times, during which they
develop some asymmetry and eventually turn over, join and deform into
a $Q=2$ single ring Hopfion.  The four remaining cases 3, 4, 5 and 6
are repulsive: As the Hopfions recede they slowly turn around their
axis and it is plausible that they eventually turn enough to attract
each other and return to form a single ring state. That is, we know
that due to the fact $2\times 1^{3/4} > 2^{3/4}$ the asymptotically
separated configuration is a higher energy state than the single-ring
configuration but it is unclear whether the initial configurations for
these four repulsive systems have large enough repulsive (escape)
energy to form the higher energy state.  As the Hopfions quickly reach
the border of any finite computational space, this is a hard question
to answer using numerical methods alone.

\subsubsection{Boundary effects}

In order to better understand the effect of boundaries, we considered
the $(++--)$ case further and calculated its evolution in physical
grids of different size, while keeping the grid constant and boundary
conditions unchanged. Our results are given in
\href{http://www.youtube.com/watch?v=tGC4_tCa1x0}{Movie W2}, which
combines the result from cubes of physical size $30.00^3$, $45.00^3$,
$67.50^3$, and $101.25^3$ having $200^3$, $300^3$, $450^3$, and
$675^3$ grid points, they are located at top left, top right, bottom
left and bottom right, respectively.

In all cases the simulation starts with a slight repulsion during
which the colors of the isosurfaces rotate towards the same phase.
Then the Hopfions approach each other and bounce several times and
eventually form a $Q=-2$ un-knot Hopfion.

\begin{figure}[f]
\begin{center}
  \includegraphics[width=12cm]{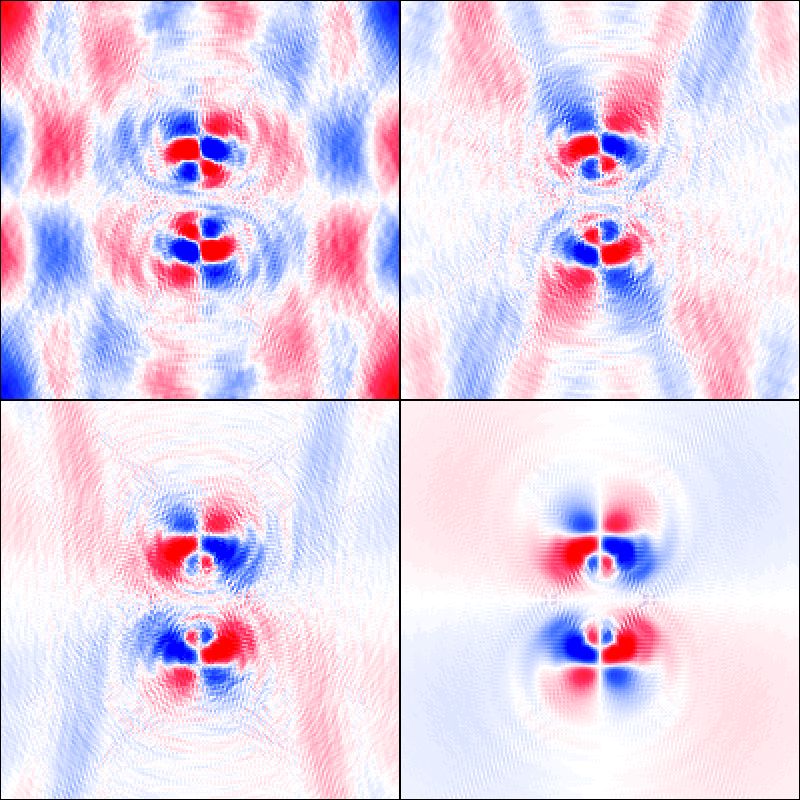}
\end{center}
\caption{Figure shows four $30^2$ slices of the canonical derivative
  field $\psi_2$ between a fixed interval $[-0.05, 0.05]$ at the same
  moment ($t=60.0$, $\#=2000$) in different physical size grids. We can
  see that the boundary effects are strongly present at the area
  around the Hopfion's core in the first case, just reaching it in the
  second and third case and still outside the observed area in the
  fourth case. \label{F:err}}
\end{figure}

The difference in the dynamics can be explained by the existence of
two competing effects: the fixing of the boundary value to
$\phi=(0,0,-1)$ tends to stabilize the configuration, while the
radiation reflecting from the boundaries tends to un-stabilize it. The
radiation effects are illustrated in Figure \ref{F:err}, where we have
plotted the values of the usually very revealing quantity $(\partial_t
\phi)$ on a slice of the box. The box in upper cases is relatively
small and the closeness of the fixed boundary stabilizes the situation
well and many bounces are required to un-stabilize it, even though
there is lots of radiation bouncing around.  The fastest
un-stabilization occurs with box size $450^3$ (lower left), in that
case radiation effects overwhelm the boundary stabilization. In the
largest box (lower right) the radiation takes a longer time to reach
the interaction region and therefore more bounces are again possible
before deformations.

Clearly the major difference between these simulation is the time
at which the radiation wave of the initial relaxation travels to the
boundary and is reflected back to center of the grid. In smaller grids
the boundary is so close that the first reflection happens before the
Hopfions even collide the first time.  After a few more reflections an
interference pattern is formed in the background but it is reasonably
symmetric.  In the larger grids the first reflected wave passes over
the Hopfion pair during the first couple of oscillations and this
asymmetric perturbation causes the state to unwind
faster.  Nevertheless the final result is always the same 

\subsubsection{Side-by-side Hopfions}
\label{Side-by-side Hopfions}
Another case considered by Ward is that of Hopfion rings side by side
on the $x$-axis. The ansatz is now
\begin{equation}\label{2solx}
W=\frac{\sigma_1(x-a)+i\sigma_2y}{z-i\tau\sigma_1\sigma_2f(r_{x-a})}+
\frac{\sigma_1'(x+a)+i\sigma_2'y}{z-i\tau\sigma_1'\sigma_2'f(r_{x+a})}.
\end{equation}
Here $r_{x\pm{}a}=((x\pm{}a)^2+y^2+z^2)^{1/2}$. In this case also the 16
possible sign combinations can be divided into 6 essentially different
ones, to which others are related by rotations:
1) $(++++)$, polarization vectors $D$ parallel,
2) $(++-+)$, antiparallel,
3) $(++--)$, parallel,
4) $(--++)$, parallel,
5) $(+++-)$, antiparallel,
6) $(+-++)$. antiparallel.
Cases 3 and 4 (and Cases 5 and 6) are related by an overall sign
change, which cannot be generated by a rotation.

Considering Hopfions asymptotically as dipole pairs and studying the
forces between them led Ward to introduce the second attractive
channel which corresponds to our Case 3 above. Ward's energy minimization
gave an axially-symmetric ring for $Q=2$. This is believed to be the
true global static minimum. \cite{Ward2000}

\begin{figure}[f]
\begin{center}
  \includegraphics[width=12cm]{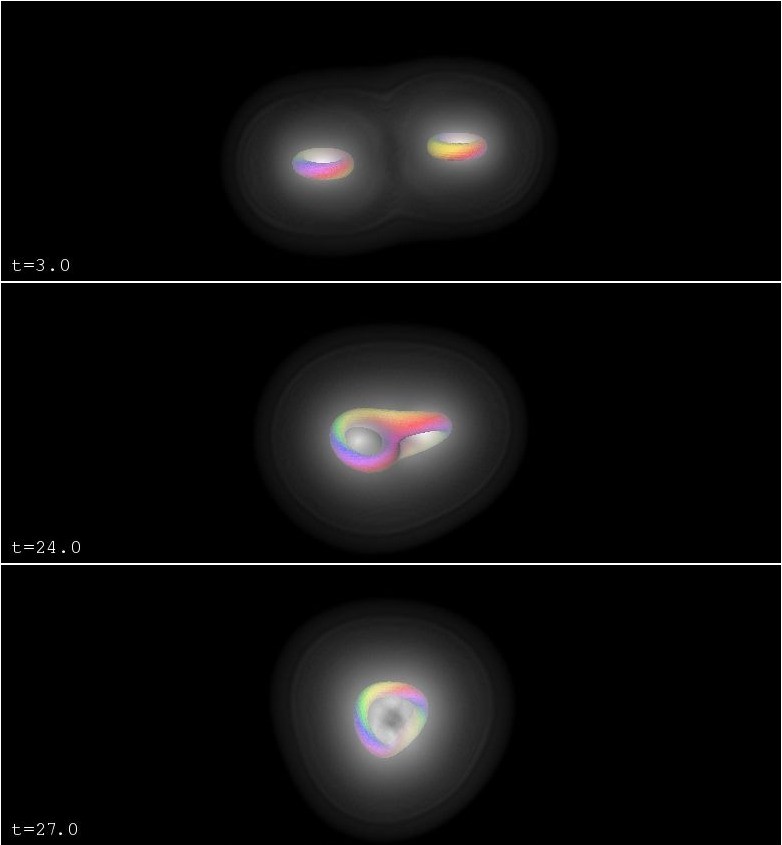}
\end{center}
\caption{The side-by-side configuration $(+++-)$ seems to be
  attractive without delay and the pair end up in the familiar $H=2$
  single ring configuration. At the end of the simulation the pair
  oscillates relatively softly having already emitted most of its
  excess energy. For more detail see
  \href{http://www.youtube.com/watch?v=fUDSIYc8O0c}{Movie
    W3}.  \label{F10}}
\end{figure}

Our numerical simulations show that the attractive and repulsive
channels yield the same qualitative behavior as in the stacked
case. However, now the attractive channel comprises of initial
configurations 3, 4, 5, and 6 while the repulsive channel contains 1
and 2. As an example of the attractive channel, we give case 5 in
Figure~\ref{F10},
\href{http://www.youtube.com/watch?v=fUDSIYc8O0c}{Movie W3}).

\begin{figure}[f]
\begin{center}
  \includegraphics[width=12cm]{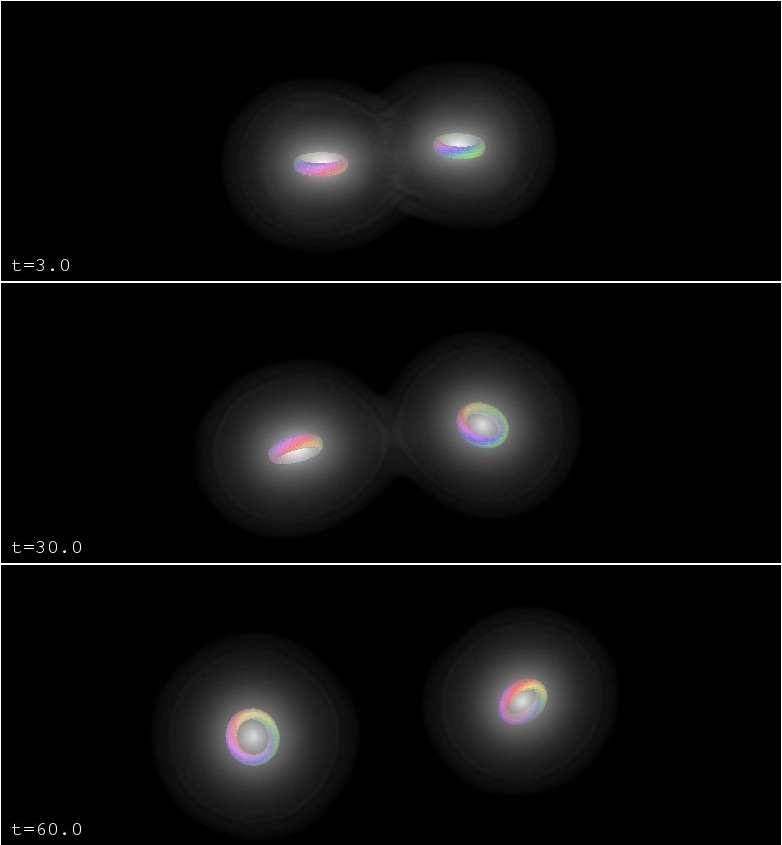}
\end{center}
\caption{The side-by-side configuration $(++-+)$ is repulsive from the
  start of the simulation and the system end up in the unattached
  state where the initial Hopfions advance in opposite directions.
\label{F11}}
\end{figure}

In order to study the repulsive case further we chose to change
somewhat the rotation angle of the initial state. After rotating the
second Hopfion by angle $\theta$ the ansatz \eqref{2solx} can be
written as
\begin{equation} \label{2solxmod}
  W = \frac{\sigma_1 (x-a)+i \sigma_2 y}{z-i \tau \sigma_1 \sigma_2 f(r_{x-a}) } 
  + \frac{\sigma'_1 \big((x+a) \cos\theta + y \sin\theta \big)+
    i \sigma'_2 \big(y \cos\theta - (x+a) \sin\theta \big)}{z-i \tau\sigma'_1
    \sigma'_2 f(r_{x+a})}.
\end{equation}
As can be seen the dependent variable $r_{x\pm{}a}$ is invariant under
rotation.

\begin{figure}[f]
\begin{center}
  \includegraphics[width=11cm]{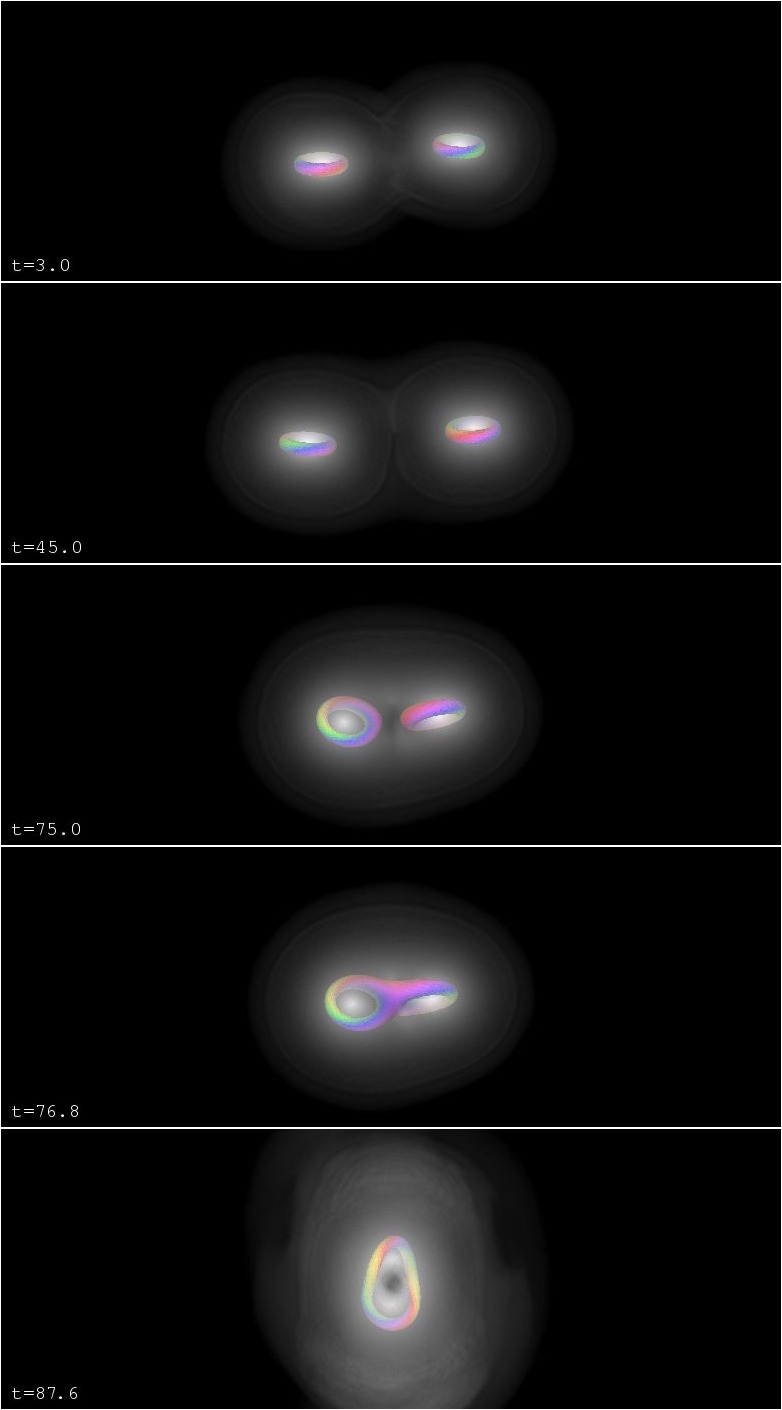}
\end{center}
\caption{The behavior of the side-by-side system $(++-+)$ changes
  after a rotation of the second Hopfion by an angle of
  $20^\circ$. Now the Hopfions move further away from each others
  under a color rotation until the phase difference between them
  becomes attractive and the single ring configuration $H=2$ is
  formed. For more detail see
  \href{http://www.youtube.com/watch?v=icjaJn21PZc}{Movie
    W4}). \label{F12}}
\end{figure}

Recall that the configurations $(+++-)$ and $(++-+)$ were respectively
in the attractive and repulsive channels. Now let us fix $\sigma$
configuration to repulsive $(++-+)$ with $\theta=0$, then the
attractive $(+++-)$ is obtained from \eqref{2solxmod} with
$\theta=180^o$. It has already been observed that even in the
repulsive case the Hopfions turn around and could return if the
computational box would be big enough. In order to study a case with
only slight repulsion we chose the configuration $(++-+)$ for
$\theta=20^\circ$ and found it to be attractive (Figure \ref{F12} and
\href{http://www.youtube.com/watch?v=icjaJn21PZc}{Movie W4}) while
$\theta=5^\circ$ was still repulsive
(\href{http://www.youtube.com/watch?v=UD_GAIUiJHg}{Movie W5}).
Similarly, for states $(++--)$ and $(++++)$: $(++++)$ is still
repulsive at $\theta=3^\circ$ but changes to attractive at
$\theta=10^\circ$.

It seems the attractive channels yield qualitatively the same behavior
regardless of whether the configuration is stacked or side-by-side and
likewise for the repulsive case. The study of the dependence on the
initial rotation angle, however, shows that there are quantitative
differences, of which we have uncovered one: the repulsive channel
contains a different range of initial relative orientations depending
on the parameters $\sigma$ and $\sigma'$. This further illustrates the
extremely rich set of phenomena associated with Hopfions.

\section{Conclusions}
Now that we are close to seeing knotted structures in experiments it
is important to understand their dynamics. The static knots in the
Faddeev-Skyrme model are already well understood and in this paper we
have shown that these objects have also dynamically many particle-like
properties.

We have here studied the scattering collisions of low charge Hopfions.
In general the results show remarkable particle-like behavior.  We
have verified again their tendency to form bound states, due to the
charge-energy dependence $E\sim |Q|^{\frac34}$; we have also seen the
associated radiation of the released energy.  With suitable speeds and
impact parameters we can also have grazing scattering, without bound
state formation. Since the internal structure of Hopfions is quite
complicated the results of the scattering process depend strongly on
the initial configuration, and not only on the initial positions and
velocities but also on the internal specifics (such as orientation,
spin, and relative phase) of the scattering objects\cite{youtube}.

We have also investigated the dynamical stability of static
configuration of two stacked or side-by-side unknots discussed by Ward
\cite{Ward2000}. As expected their evolution depends sensitively on
computational aspects due to the mere length of the calculation (due
to the initial state being static). The errors arise both from
discretizations and in particular from boundary effects of the finite
computational box. It seems that the stacked and side-by-side
configurations made of two charge 1 Hopfions are all unstable and
eventually relax into a charge 2 Hopfion.

\section*{Acknowledgments}
This work is supported by Academy of Finland grants 47188 and 123311,
and by the UK Engineering and Physical Sciences Research Council. The
computations were done using the resources provided by the {\em CSC --
  IT Center for Science Ltd}, {\tt www.csc.fi} and by the {\em M-grid
  project}, supported by the Academy of Finland, and by {\em CSC -- IT
  Center for Science Ltd}.

\end{document}